\documentclass[11pt,letterpaper]{article}
\pdfoutput=1
\usepackage{jheppub}
\usepackage{bbm}
\usepackage{mathrsfs}
\usepackage{slashed}
\usepackage{caption}
\usepackage{epstopdf}
\usepackage[normalem]{ulem}
\usepackage[bottom]{footmisc}
\usepackage{subcaption}
\usepackage{bbold}
\usepackage{titlesec}
\usepackage{threeparttable}
\usepackage{booktabs}
\usepackage{changepage}
\usepackage[utf8]{inputenc}
\usepackage{dsfont} 

\usepackage{grffile}
 
\usepackage{graphicx}   
\usepackage{dcolumn}   
\usepackage{bm}        
\usepackage{amssymb}   
\usepackage{setspace}
\usepackage{amsmath, amssymb, setspace}
\usepackage{array}
\usepackage{booktabs}
\usepackage{caption}
\usepackage{indentfirst}
\usepackage{float}
\usepackage{lmodern}
\usepackage{multirow}
\usepackage{soul}
\usepackage[normalem]{ulem}

\usepackage{braket}
\usepackage{comment}

\usepackage[draft]{pgf}

\usepackage{adjustbox} 
\newlength{\myimageoversize}
\newsavebox{\myimage}

\usepackage[most]{tcolorbox}

\usepackage{empheq}

\tcbset{colback=white!10!white, colframe=black!50!black, 
        highlight math style= {enhanced, 
            colframe=black,colback=white!10!white,boxsep=0pt}
        }

\titleformat{\subsubsection}
{\normalfont\fontsize{12}{17}\itshape}{\thesubsubsection}{1em}{}
 
\title{\Huge{Interstellar Gas Heating by Primordial Black Holes}}

\author[a,b]{Volodymyr Takhistov,}
\author[b]{Philip Lu,}
\author[b]{Graciela B. Gelmini,}
\author[c,d]{Kohei Hayashi,}
\author[e,f,a]{Yoshiyuki Inoue,}
\author[b,a]{Alexander Kusenko}
\affiliation[a]{Kavli Institute for the Physics and Mathematics of the Universe (WPI), University of Tokyo, Kashiwa 277-8583, Japan}
\affiliation[b]{Department of Physics and Astronomy, University of California, Los Angeles, CA 90095-1547, USA}
\affiliation[c]{Astronomical Institute, Tohoku University, Sendai 980-8582, Japan}
\affiliation[d]{National Institute of Technology, Ichinoseki College, Ichinoseki 021-8511, Japan}
\affiliation[e]{Department of Earth and Space Science, Graduate School of Science, Osaka University, Toyonaka, Osaka 560-0043, Japan}
\affiliation[f]{Interdisciplinary Theoretical \& Mathematical Science Program (iTHEMS), RIKEN, 2-1 Hirosawa, Saitama 351-0198, Japan} 

\abstract{
Interstellar gas heating
is a powerful cosmology-independent observable for exploring the parameter space of primordial black holes (PBHs) formed in the early Universe that could constitute part of the dark matter (DM). We provide a detailed analysis of the various aspects for this observable, such as PBH emission mechanisms. Using observational data from the Leo T dwarf galaxy, we constrain the PBH abundance over a broad mass-range, $M_{\rm PBH} \sim \mathcal{O}(1) M_{\odot}-10^7 M_{\odot}$, relevant for the recently detected gravitational wave signals from intermediate-mass BHs. We also consider PBH gas heating of systems with bulk relative velocity with respect to the DM, such as Galactic clouds.}

\emailAdd{volodymyr.takhistov@ipmu.jp}
\emailAdd{philiplu11@gmail.com}
\emailAdd{gelmini@physics.ucla.edu}
\emailAdd{kohei.hayashi@g.ichinoseki.ac.jp}
\emailAdd{yinoue@astro-osaka.jp}
\emailAdd{kusenko@ucla.edu}

\begin{document}

\preprint{IPMU21-0029\\
\rightline{RIKEN-iTHEMS-Report-21}}

 \maketitle
\flushbottom

\section{Introduction}
\label{sec:intro}

Primordial black holes (PBHs) can form in the early Universe, prior to galaxies and stars, through a variety of mechanisms and can constitute a significant fraction of the dark matter (DM)~(e.g.~\cite{Zeldovich:1967,Hawking:1971ei,Carr:1974nx,Khlopov:1980mg,Khlopov:1985jw,Kawasaki:1997ju,GarciaBellido:1996qt,Khlopov:2008qy,Frampton:2010sw,Kawasaki:2016pql,Carr:2016drx,Inomata:2016rbd,Pi:2017gih,Inomata:2017okj,Garcia-Bellido:2017aan,Georg:2017mqk,Inomata:2017vxo,Kocsis:2017yty,Ando:2017veq,Cotner:2016cvr,Cotner:2017tir,Cotner:2018vug,Sasaki:2018dmp,Carr:2018rid,Banik:2018tyb,1939PCPS...35..405H,Cotner:2019ykd,Kusenko:2020pcg,Flores:2020drq,Domenech:2021uyx}). Depending on the formation mechanism, the PBH mass can span many orders of magnitude. Those PBHs with mass above $M \gtrsim 10^{15}$~g survive Hawking evaporation until present. In a sizable mass range, $M \sim 10^{-16} - 10^{-10} M_{\odot}$, PBHs can constitute all of the DM~\cite{Katz:2018zrn,Smyth:2019whb,Montero-Camacho:2019jte,Sugiyama:2019dgt,Niikura:2017zjd}. PBHs have been associated with a variety of astrophysical phenomena and puzzles, e.g. PBHs could be the seeds for the formation of super-massive black holes~(e.g.~\cite{Bean:2002kx,Kawasaki:2012kn,Clesse:2015wea}) and could lead to new signatures~from compact star disruptions from PBH capture~(e.g.~\cite{Fuller:2017uyd,Takhistov:2017nmt,Takhistov:2017bpt,Takhistov:2020vxs}).  

PBHs in the mass window of $\sim 10-10^3 M_\odot$ have been suggested as progenitors \cite{Nakamura:1997sm,Clesse:2015wea,Bird:2016dcv,Raidal:2017mfl,Eroshenko:2016hmn,Sasaki:2016jop,Clesse:2016ajp} for the gravitational wave (GW) signals detected by LIGO~(e.g. \cite{Abbott:2016blz,Abbott:2016nmj,Abbott:2017vtc}). The first definitive observation of intermediate-mass $150 M_\odot$ BH has been claimed by LIGO \citep{Abbott:2020tfl}. In conventional stellar evolution astrophysics such BHs are difficult to form from the collapse of a single star, akin to the formation of stellar-mass BHs. While this PBH DM parameter space region is already under pressure from a variety of constraints
(see Ref.~\citep{Carr:2020gox} for review), these rely on and are sensitive to uncertainties of underlying assumptions, which often include the cosmological history.

In Ref.~\cite{Lu:2020bmd} we have put forth a novel cosmology-independent observable, based on PBHs heating of the surrounding interstellar medium (ISM) gas, as a powerful probe for stellar and intermediate-mass PBHs contributing to the DM. In this work we further explore this observable and provide a detailed analysis of PBH gas heating as well as various underlying processes, such as PBH emission mechanisms. Following Ref.~\cite{Lu:2020bmd}, we focus on the Leo T irregular dwarf galaxy as a particularly favorable astrophysical system target. Further, we extend the analysis to include systems with bulk velocity relative to the DM.

The ``heating'' of near-by stars by passing massive BHs has been previously considered~\cite{Carr:1999dyn,Lacey:1985,Totani:2010}. Theoretically expected X-ray emission from traversing PBHs interacting with the surrounding medium has been also employed to set limits on the PBH DM fraction~\cite{Gaggero:2016dpq,Inoue:2017csr,Hektor:2018rul}. Constraints from gas heating and system cooling have been considered in the context of particle DM candidates~\cite{Bhoonah:2018wmw,Farrar:2019qrv,Wadekar:2019mpc}, which however have heating mechanisms different from PBHs. Further, while the focus for particle DM is on target astrophysical systems with high relative particle velocities resulting in high collision rates, for PBHs the astrophysical systems with slower relative velocities are favored due to increased PBH accretion and resulting emission. 

While our discussion focuses on present-day effects, emission from gas-accreting PBHs is also relevant for cosmology, e.g. for the reionization and recombination epochs (e.g. \cite{Ricotti:2007au}). X-ray emission from gas accreting onto PBHs would result in ionization and increased quantities of molecular hydrogen, which would in turn provide cooling and enhance earlier star formation~\cite{Ricotti:2007au}. Further, emission from accreting PBHs could result in spectral $\mu-$ and $y-$distortions of the cosmic microwave background (CMB).
Heating and ionization of intergalactic medium gas from PBH emission could modify 21cm power spectrum in an observable way~\cite{Hektor:2018qqw,Mena:2019nhm}. 
In this work we do not discuss cosmological effects, which provide complementary ways of exploring PBH DM.  

\section{PBHs in ISM}
\label{sec:accretion}

\subsection{Bondi-Hoyle-Lyttleton accretion}
\label{ssec:bondihoyle}

X-ray emission from gas-accreting isolated BHs traversing the ISM of the Milky Way was explored in~Ref.~\cite{Agol:2001hb}. Ref.~\cite{Inoue:2017csr} adapted the analysis for PBHs constituting part of the DM. In this section we broadly follow the same method. PBHs traveling at low Mach number $\mathcal{M} = v/c_s$ accrete via the Bondi-Hoyle-Lyttleton process, with an accretion rate~\cite{1939PCPS...35..405H,1944MNRAS.104..273B,1952MNRAS.112..195B}
\begin{align}
\label{eq:bondihoyle}
    \dot{M} = 4 \pi r_B^2 \tilde{v} \rho = \frac{4\pi G^2 M^2 n \mu m_p}{\Tilde{v}^3}~.
\end{align}
Here $M$ is the PBH mass, $r_B = G M/\tilde{v}^2$ is the Bondi radius\footnote{Bondi radius sets
the scale at which gravity becomes important and approximates how far from object material starts to be drawn in and accreted.}, $\mu$ is the mean molecular weight, $n$ is the ISM gas number density, $m_p$ is the proton mass and
 $\tilde{v} \equiv (v^2+c_s^2)^{1/2}$, where $v$ is the PBH velocity relative to the ISM gas and $c_s$ is the  temperature-dependent sound speed in gas, which we take to be approximately $c_s \sim 10$~km/s~\cite{Inoue:2017csr}. At high Mach number, the BH accretion rate is expected to be limited to $\sim10$--$20$\% of the canonical Bondi-Hoyle-Lyttleton accretion rate of Eq.~\eqref{eq:bondihoyle}~\cite{Guo:2020qkf}, as shown by recent 3D hydrodynamical simulations.
 However, the resulting system behavior is sensitive to assumed initial conditions. We find that it is the low-velocity regime that predominantly contributes to our effects of interest. Hence, we adopt the standard accretion rate of Eq.~\eqref{eq:bondihoyle}.

The bolometric luminosity of the accretion disk of a BH can be related to its accretion rate as $L=\epsilon(\dot{M})\dot{M}$. Here $\epsilon(\dot{M})$ is an effective phenomenological parameter describing radiative efficiency, which varies according to different accretion regimes.
Assuming a radiative efficiency of $\epsilon_0 = 0.1$ characteristic of thin disks\footnote{The radiative efficiency $\epsilon_0$ can vary from 0.057 for a non-rotating Schwarzschild BH to
0.42 for an extremal Kerr BH~(see e.g.~Ref.~\cite{Kato:book2008}).}, the Eddington accretion rate in terms of Eddington luminosity is
\begin{equation}
\label{eq:eddingtonrate}
    \dot{M}_\textrm{Edd} = \frac{L_\textrm{Edd}}{\epsilon_0 c^2} = 6.7\times10^{-16}\left(\frac{M}{M_\odot}\right)~M_\odot\text{/s}~.
\end{equation}
To determine the accretion regimes of interest, it is useful to introduce a dimensionless form of the accretion rate
\begin{equation}
\label{mdot}
    \dot{m} = \frac{\dot{M}}{\dot{M}_\textrm{Edd}} = 2.64\times10^{-7} \left(\frac{M}{M_\odot}\right)\left(\frac{n}{1\textrm{ cm}^{-3}}\right) \left(\dfrac{\mu}{1}\right) \left(\frac{\Tilde{v}}{10\textrm{ km}/\textrm{s}}\right)^{-3}~. 
\end{equation} 

\subsection{Accretion disk formation}

An accretion disk can form around a BH from the gas infalling with sufficient angular momentum.
For accretion rates $\dot{m}\gtrsim 0.07\alpha$, a standard thin $\alpha$-disk~\cite{Shakura:1972te} likely forms. Here the disk's viscosity is described by a phenomenological parameter $\alpha \sim \mathcal{O}(0.1)$.
Perturbations in the density or velocity distributions of the surrounding gas can supply angular momentum to the disk. Assuming a Kolmogorov spectrum of density fluctuations in the ISM, the resulting density differential on opposing sides of the accretion ``cylinder'' can be calculated~\cite{1976ApJ...204..555S}. We equate the angular momentum from this differential with the angular momentum of the outer edge of the thin disk~\cite{Agol:2001hb},
\begin{equation}
\label{eq:router}
    r_{\rm out} \simeq 5.4\times10^{9} r_s \left(\frac{M}{100 M_\odot}\right)^{\frac{2}{3}}\left(\frac{\Tilde{v}}{10 \textrm{ km/s}}\right)^{-\frac{10}{3}}~,
\end{equation}
where $r_s = 2 G M/c^2$ is the BH Schwarzschild radius. On the other hand, for inefficient accretion described by the Advection-Dominated Accretion Flow (ADAF) regime~\cite{Narayan:1994xi,Yuan:2014gma}, the inflowing matter does not follow a Keplerian description. In this case, the outer accretion disk radius is uncertain and we consider a range of parameter values as described in Sec.~\ref{sec:heating}.  

The inner edge of the accretion disk is assumed to follow
the innermost stable circular orbit (ISCO) of a test particle, which for a Schwarzschild BH is $r_{\rm ISCO} = 3 r_s$. If the outer edge of the BH accretion disk, at $r_{\rm out}$, is found to be located within the ISCO radius, i.e. $r_{\rm out}/r_{\rm ISCO} < 1$, we do not expect efficient disk formation.
Thus, using Eq.~\eqref{eq:router} we can infer an approximate condition on the mass of a BH for accretion disk formation as
\begin{equation}
M > 2.5 \times 10^{-13} M_{\odot} \left(\frac{\Tilde{v}}{10 \textrm{ km/s}}\right)^{5}~.
\end{equation}
This condition is well-satisfied for all BHs in our parameter range of interest. On the other hand, it shows that small PBHs are generally not expected to have an accretion disk.

\subsection{Gas and PBH distributions}
\label{ssec:pbhdist}

The total number of PBHs of mass $M$ within a gas system of volume $V$ and DM density $\rho_{\rm DM}$, assuming PBHs make up a fraction $f_{\rm PBH}$ of the DM, is given by 
\begin{equation}
N_{\rm PBH}(M) = f_{\rm PBH} \frac{\rho_{\rm DM}V}{M}~.
\end{equation}
Our considerations apply in case there is at least one PBH in the gas system. This results in a lower ``incredulity" limit
\begin{equation}
\label{eq:incredlimit}
    f_\textrm{bound} > \frac{3M}{4\pi r_\textrm{sys}^3 \rho_\textrm{DM}}~,
\end{equation}
where $r_\textrm{sys}$ is the size of the system\footnote{For simplicity, we assume spherical DM halo distribution as convention, however other distributions are also possible~(e.g.~\cite{Hayashi:2012si})}.

Here we assume a monochromatic PBH mass function, however our results can be readily applied to an extended PBH mass distribution. The velocity of PBHs contributing to the DM can be described by a Maxwell-Boltzmann distribution 
\begin{equation}
\label{eq:maxwellian}
    \frac{df_v}{dv}(v) = \sqrt{\frac{2}{\pi}}\frac{v^2}{\sigma_v^3}\exp\left(-\frac{v^2}{2\sigma_v^2}\right)~,
\end{equation}
where $\sigma_v$ is the velocity dispersion within a given system.

We are interested in the gas heating due to PBHs. For a gas system in thermal equilibrium, we estimate the total amount of heating $H_{\rm tot}$ from PBHs of mass $M$ as
\begin{equation}
\label{eq:totalheateq}
    H_{\rm tot}(M) = N_{\rm PBH}(M) H(M) \simeq N_{\rm PBH}(M)\int_{n_\textrm{min}}^{n_\textrm{max}}\int_{v_\textrm{min}}^{v_\textrm{max}} dn dv \frac{df_n}{dn}\frac{df_v}{dv}\mathcal{H}(M,n,v)~,
\end{equation}
where $df_n/dn$ is the gas number density distribution, $df_v/dv$ is the relative PBH velocity distribution and $\mathcal{H}(M,n,v)$ is the amount of heat being deposited into the system from a single PBH and also takes into account other potential contributing effects (e.g. photon absorption). $\mathcal{H}$ represents the cumulative contribution of all three heating mechanisms we consider below. For photon emission as well as outflows, there is an additional integration to treat absorption. For the gas systems of interest the density is approximately constant, for which $df_n/dn$ can be taken as a delta function in $n$.

\section{Gas Heating Mechanisms}
\label{sec:heating}

We consider three distinct gas heating mechanisms due to PBHs: absorption of photons emitted from the accretion disk, direct heating due to dynamical friction and energy deposit by protons of outflows that are inherent to the ADAF regime. Of these three, dynamical friction has the least uncertainty but typically constitutes a subdominant contribution. Since for our relevant parameters of interest the PBHs are in the stellar and intermediate-mass BH-range, the heating associated with Hawking evaporation is negligible. Following the original proposal of Ref.~\cite{Lu:2020bmd}, PBH gas heating has been recently considered in the context of small evaporating PBHs as well~\cite{Kim:2020ngi,Laha:2020vhg}.

\subsection{Accretion photon emission}
\label{ssec:xrays}

Photon emission from PBH accretion disks is one of the major contributors to the ISM gas heating. Emission in the X-ray band is most readily absorbed by the surrounding hydrogen ISM gas, becoming efficient when the mass accretion rates are significant.

\begin{figure}[tb]
\begin{center}
\includegraphics[trim={5mm 0mm 40 0},clip,width=0.65\textwidth]{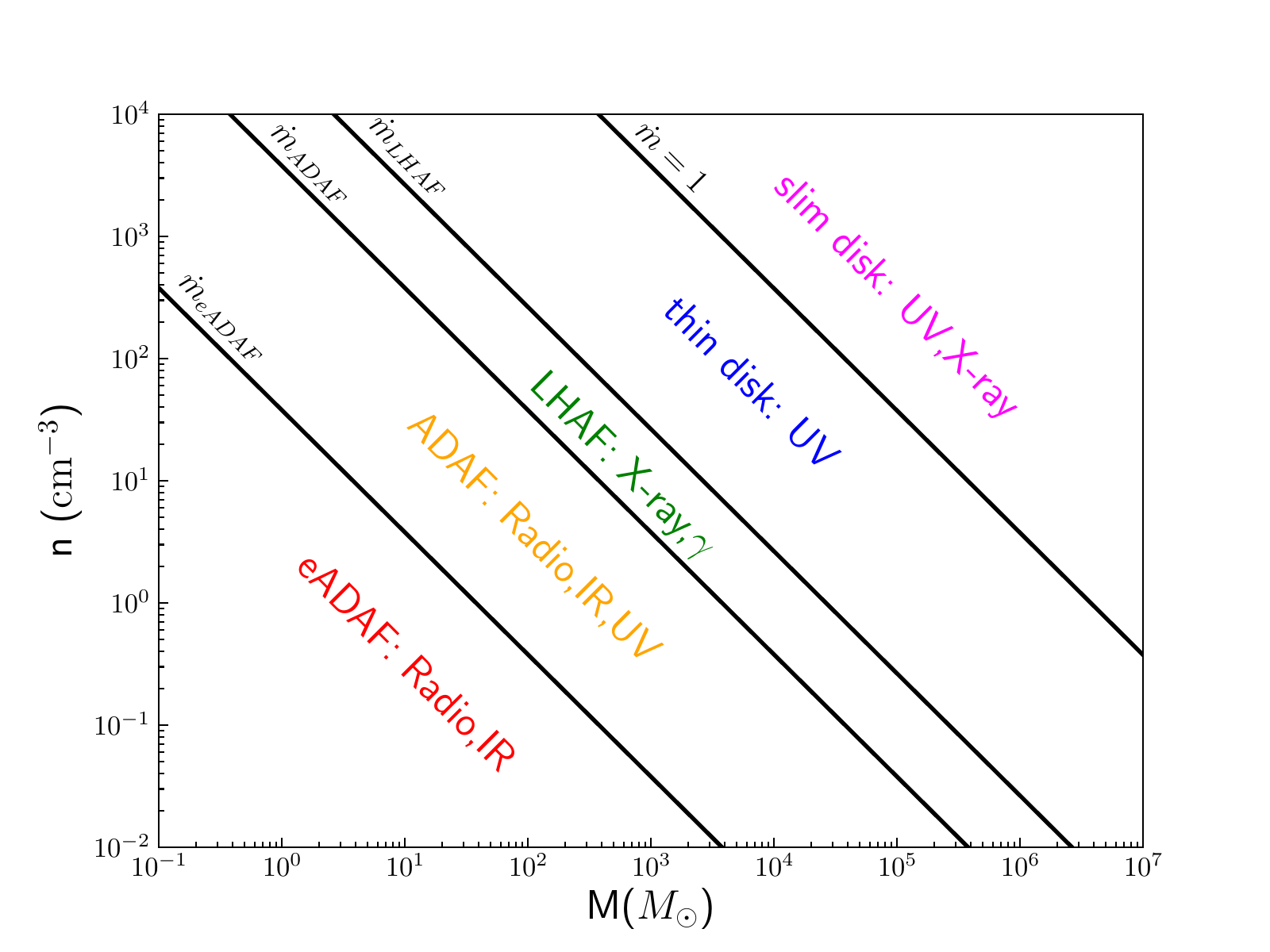}

\caption{  \label{fig:emissionregions} 
Dominant photon emission from PBH accretion over a wide mass-range of PBHs $M$, for different number densities $n$ of the surrounding gas, assuming Bondi-Hoyle-Lyttleton accretion and a characteristic flow velocity of $\tilde{v} \simeq 10$~km/s in Eq.~\eqref{eq:bondihoyle}. Regions of different accretion regimes $\dot{m}$ (black diagonal lines) resulting in slim disk, thin disk and ADAF (including LHAF, ADAF, eADAF sub-regimes, see text) accretion flows are shown.}
\end{center}
\end{figure}

For different accretion disk regimes we follow Ref.~\cite{Yuan:2014gma}.  We consider that the accretion flow results in a thin disk formation when $\dot{m}>\dot{M}=0.07\alpha$, and is described by ADAF for less efficient accretion. We do not consider the ``slim disk'' and other solutions relevant for near-Eddington or super-Eddington accretion when $\dot{m}\sim 1$, as the PBHs in our parameter space of interest are not expected to typically achieve such high rates. Furthermore, the slim disk emission spectrum is not significantly different from the spectrum of the thin disk for $\dot{m}\lesssim 10$~\cite{1999ApJ...522..839W}. In Fig.~\ref{fig:emissionregions}, we display 
the primary photon emission band from accreting PBHs over a wide range of PBH masses and for different densities of the surrounding gas, assuming Bondi-Hoyle-Lyttleton accretion of Eq.~\ref{eq:bondihoyle} with a characteristic flow velocity of $\tilde{v} \simeq 10$~km/s, and emission results of Sec.~\ref{sssec:thindisk} and Sec.~\ref{sssec:ADAF}. Fig.~\ref{fig:emissionregions} suggests which PBH observational signatures could be generically expected for a particular environment. In certain accretion regimes the disk size could grow to become comparable to the Bondi radius and feedback effects on the accretion can appear. However, we expect that PBHs can still efficiently accrete, e.g. from outside the plane of the disk. Throughout we assume that Bondi accretion continues to hold.

Hydrogen gas absorbs ionizing radiation efficiently, but below the threshold of $E_i = 13.6~\textrm{eV}$, hydrogen is optically thin to continuous emission. For non-relativistic gas velocity dispersions, Doppler broadening of the emission spectra is not significant so that we can ignore scattering with photons that have energies lower than $E_i$. The majority of photons are absorbed and a significant fraction of the energy is deposited as heat when the medium is optically thick. For $13.6\textrm{ eV}<E<30\textrm{ eV}$, we employ the photo-ionization cross-section of Refs.~\cite{Bethe1957,1990A&A...237..267B},
\begin{equation}
\label{eq:ionizationformula}
    \sigma(E) = \sigma_0 y^{-\frac{3}{2}}\left(1+y^{\frac{1}{2}}\right)^{-4}~,
\end{equation}
where $y=E/E_0$, $E_0 = 1/2 E_i$ and $\sigma_0 = 605.73 \textrm{ Mb} = 6.06\times10^{-16}\textrm{ cm}^2$. The optical depth for a gas system of size $l$ and number density $n$ is then given by 
\begin{equation}
    \tau (n, E) = \sigma(E) n l~.
\end{equation}
Above $30\textrm{ eV}$, we use attenuation length data from Fig.~(32.16) of Ref.~\cite{Olive_2014}. 

The resulting heating power due to accretion disk emission photons is
\begin{equation}
\label{eq:heatingphotons}
    \mathcal{H}(M,n,v) = \int_{E_i}^{E_\textrm{max}}L_\nu (M,n,v) f_h \left(1-e^{-\tau}\right)d\nu~,
\end{equation}
where $L_{\nu}(\nu)$ is the photon luminosity at frequency $\nu$ and $f_h$ is the fraction of energy deposited as heat that we estimate to be $\sim 1/3$~\cite{1985ApJ...298..268S,Ricotti:2001zf,Furlanetto:2009uf,Kim:2020ngi}.
Both the ADAF and thin disk emission spectra decrease exponentially at high energies, so we evaluate the integral up to a maximum energy $E_{\rm max}$, above which the contributions are negligible\footnote{For thin disk, we choose the maximum temperature to be $E_{\rm max} = 5 T_i$, where $T_i$ is the characteristic temperature of the inner disk. For ADAF, we take $E_{\rm max} = 3 T_e$, with $T_e$ being the temperature of the electron plasma.}.

\subsubsection{Thin disks}
\label{sssec:thindisk}

The standard (geometrically) thin  $\alpha$-disk, characterized by the viscosity parameter $\alpha$, is optically thick and efficiently emits blackbody radiation~\cite{Shakura:1972te}. Thin disk allows for a fully analytic description. The temperature of the disk varies with radius as~\cite{Pringle:1981ds}
\begin{equation}
\label{eq:alphatemp}
    T(r) = T_i \left(\frac{r_i}{r}\right)^{3/4} \left[1-\left(\frac{r_i}{r}\right)^{\frac{1}{2}}\right]^{\frac{1}{4}}~,
\end{equation}
where $r_i$ is the inner disk radius taken to be the ISCO radius $r_{\rm ISCO} = 3 r_s$ as before. Here, using Eq.~\eqref{eq:bondihoyle},
\begin{equation} \label{eq:tialpha}
    T_i = \left(\frac{3GM\dot{M}}{8\pi r_i^3 \sigma_B}\right)^{1/4} = 53.3\textrm{ eV}\left(\frac{n}{1\textrm{ cm}^{-3}}\right)^{\frac{1}{4}} \left(\frac{\Tilde{v}}{10\textrm{ km}/\textrm{s}}\right)^{-\frac{3}{4}}~,
\end{equation}
where $\sigma_B$ is the Stefan-Boltzmann constant.
For Bondi-Hoyle-Lyttleton accretion, the disk temperature is seen to be independent of the BH mass. The disk temperature $T(r)$ decreases at large radii as $r^{-3/4}$, but also diminishes near the inner radius. The disk attains a maximum temperature of $T_{\rm max} = 0.488 T_i$ at a radius of $r = 1.36 r_i$.

The thin disk emission spectrum is a combined contribution of the blackbody spectra from each radius. Using the scaling relations in Ref.~\cite{Pringle:1981ds} and requiring continuity, the
resulting spectrum can be approximately described as
\begin{align}
\begin{split}
\label{eq:thinspectrum}
    \nu < T_o :~~~~~& L_\nu = c_\alpha \left(\frac{T_{\rm max}}{T_o}\right)^{\frac{5}{3}}\left(\frac{\nu}{T_\textrm{max}}\right)^2 \\
    T_o < \nu < T_{\rm max} :~~~~~& L_\nu = c_\alpha \left(\frac{\nu}{T_\textrm{max}}\right)^{\frac{1}{3}} \\ 
    T_{\rm max} < \nu :~~~~~& L_\nu = c_\alpha \left(\frac{\nu}{T_\textrm{max}}\right)^2 e^{1-\frac{\nu}{T_{\rm max}}}
\end{split}    
\end{align}
where 
\begin{equation}
c_\alpha = 1.27\times10^{29}\textrm{ erg}\textrm{ eV}^{-1}\textrm{s}^{-1} \Big(\dfrac{M}{M_\odot}\Big)^2 \Big(\dfrac{n}{1\textrm{ cm}^{-3}}\Big)^{3/4}\Big(\dfrac{\Tilde{v}}{10\textrm{ km}/\textrm{s}}\Big)^{-9/4}~.
\end{equation}
Here $c_\alpha$ is normalized so that the total integrated luminosity (power) is $\int L_{\nu} d\nu = 0.057\dot{M}c^2$, i.e.~emission has the maximum possible efficiency for a Schwarzschild black hole. Using Eq.~\eqref{eq:bondihoyle}, \eqref{eq:router} and \eqref{eq:alphatemp}, the temperature at the outer disk radius $T_o = T(r_{\rm out})$ is given by
\begin{equation}
\label{eq:touter}
    T_o \simeq 6.13\times10^{-5}\textrm{ eV} \left(\frac{n}{1\textrm{ cm}^{-3}}\right)^{\frac{1}{4}}\left(\frac{M}{M_\odot}\right)^{-\frac{1}{2}}\left(\frac{\Tilde{v}}{10\textrm{ km}/\textrm{s}}\right)^{\frac{7}{4}}~.
\end{equation}
For all relevant parameter values of interest for $n,M,\Tilde{v}$, the temperature is always $T_o \ll 13.6~\textrm{eV}$.

Eq.~\eqref{eq:tialpha} shows that thin disks efficiently emit at X-ray frequencies, a regime where the hydrogen ionization cross section is large. Hence, the emission from thin disks would be well absorbed and could contribute significantly to gas heating.

\subsubsection{ADAF - emission power}
\label{sssec:ADAF} 

When the accretion is significantly sub-Eddington, ADAF disks form~\cite{Narayan:1994xi,Yuan:2014gma}.
Here, the heat generated by viscosity during accretion is not efficiently radiated out and much of the energy is advected via matter heat capture into the BH event horizon, along with the gas inflow. In contrast to the thin disk, the ADAF disk is geometrically thick and optically thin. 

An ADAF disk has a complex multi-component emission spectrum, of which we will consider the three dominant ones associated with electron cooling\footnote{We neglect additional possible contribution of synchrotron radiation from non-thermal electrons, which is present in ADAF models of Sgr A$^\ast$~\cite{2003ApJ...598..301Y}.}: synchrotron radiation, inverse Compton (IC) scattering and bremsstrahlung. 

To describe the ADAF spectrum, we will employ approximate analytic expressions obtained in Ref.~\cite{Mahadevan:1996jf} in combination with the updated values for the phenomenological input parameters. The accretion inflow forms a hot two-temperature plasma~\cite{1995ApJ...452..710N}, with the hotter ions at $T_\textrm{ion}\simeq 10^{11} K$ and the cooler radiating electrons with $T_e \simeq 10^{9}-10^{10} K$. We consider the temperature to be spatially uniform over the entire emission region. To characterize the flow, we use the following parameter values consistent with recent numerical simulations and observations~\cite{Yuan:2014gma}: the fraction of the viscously dissipated energy that directly heats
electrons\footnote{The remainder $(1-\delta)$ goes into the ions.} $\delta = 0.3$, the ratio of gas pressure to total pressure $\beta = 10/11$, the minimum flow radius is the ISCO radius with $r_\textrm{min}=3r_s$, and the viscosity parameter $\alpha = 0.1$. For our results below, we employ $x_M \simeq 4\times 10^3 \dot{m}^{1/4}$ and constants $c_1=0.5$, $c_3=0.3$ in the formalism of Ref.~\cite{Mahadevan:1996jf} to characterize the ADAF solution~\cite{Narayan:1994xi}. 

For a more robust analysis, we will further divide ADAF into three distinct sub-regimes, depending on the accretion rate, following Ref.~\cite{Yuan:2014gma}.  Namely, we shall consider luminous hot accretion flow (LHAF) for accretion rates $\dot{M}_\textrm{ADAF}=0.1\alpha^2 < \dot{m} < \dot{M}_\textrm{LHAF} =0.07\alpha $, standard ADAF in the range $10^{-3} \alpha^2=\dot{M}_\textrm{eADAF} < \dot{m} < \dot{M}_\textrm{ADAF}$, and ``electron" ADAF (eADAF) for $\dot{m} < \dot{M}_\textrm{eADAF}$. As discussed below in Section~\ref{sssec:ADAFtemp}, the temperature determination is handled differently for each of these ADAF regimes.

The synchrotron emission is self-absorbed and produces a rising $L_\nu \propto \nu^{5/2}$ spectrum, similar to the $\nu^{1/3}$ spectrum in the optically thin regime and also to that of the thin disk, which peaks at
\begin{equation}
\label{eq:nup}
    \nu_p =1.83\times10^{-2}\textrm{ eV} \left(\frac{\alpha}{0.1}\right)^{-\frac{1}{2}}\left(\frac{1-\beta}{1/11}\right)^{\frac{1}{2}}\theta_e^2\left(\frac{r_\textrm{min}}{3 r_s}\right)^{-\frac{5}{4}}\left(\frac{M}{M_\odot}\right)^{-\frac{1}{2}}\left(\frac{\dot{m}}{10^{-8}}\right)^\frac{3}{4}~,
\end{equation}
where
\begin{equation}
\theta_e=\frac{kT_e}{m_ec^2}=\frac{T_e}{5.93\times10^9 K}
\end{equation}
is the temperature in units of the electron mass $m_e$. The peak luminosity is given by
\begin{equation}
\label{eq:lnup}
    L_{\nu_p} = 5.06\times10^{38}\frac{\textrm{erg}}{\textrm{s}\cdot\textrm{eV}}~ \alpha^{-1}(1-\beta)\left(\frac{M}{M_\odot}\right)\dot{m}^\frac{3}{2}\theta_e^5r_\textrm{min}^{-\frac{1}{2}}~.
\end{equation}
The total resulting synchrotron power is
\begin{equation}
P_\textrm{syn}= \int_0^{\nu_p} L_{\nu} d\nu \simeq 0.71\nu_p L_{\nu_p}~.
\end{equation}
For our typical parameters of interest, the synchrotron emission is poorly absorbed by the surrounding medium, since it peaks well below the hydrogen ionization threshold. Therefore, it does not significantly contribute to gas heating.

The synchrotron photons undergo inverse-Compton scattering with the surrounding electron plasma. The thermal IC spectrum can be well approximated by a power law in the frequency range $\nu_p\leq\nu\lesssim3 k T_e$ as
\begin{equation} 
\label{eq:ICspectrum}
L_{\nu,\textrm{IC}} = L_{\nu_p}\left(\frac{\nu}{\nu_p}\right)^{-\alpha_c}~,
\end{equation}
where
 \begin{equation} \label{eq:alphac}
    \alpha_c = -\frac{\ln{\tau_{es}}}{\ln{A}}~.
\end{equation}
The exponent $\alpha_c$ depends on the electron scattering optical depth $\tau_{es}=12.4\dot{m}\alpha^{-1}r_\textrm{min}^{-1/2}$ and the amplification factor (mean amplification per scattering) $A=1+4\theta_e+16\theta_e^2$. The total IC power
\begin{equation}
\label{eq:ICpower}
    P_\textrm{IC}=\int_{\nu_p}^{3T_e}L_{\nu,\textrm{IC}}d\nu = \dfrac{\nu_p L_{\nu_p}}{1-\alpha_c}\left(\left[\frac{6.25\times 10^{7} \left(\dfrac{T_e}{10^{9}\textrm{ K}}\right)}{\left(\dfrac{\nu_p}{10^{12}}\right)}\right]^{1-\alpha_c}-1\right)
\end{equation}
is sensitive to the considered temperature due to the $T^5$ dependence of $L_{\nu_p}$ and the temperature dependence implicit in $\alpha_c$. 

Depending on the accretion rate, different components of the ADAF spectrum can dominate. At low mass accretion rates (eADAF regime), the exponent in IC luminosity becomes $\alpha_c>2$, resulting in $P_\textrm{IC}\simeq P_\textrm{syn}/0.71(\alpha_c-1)$, and the IC power is a fraction of the synchrotron power. For moderate accretion rate (ADAF regime), $1<\alpha_c\lesssim 2$ and while IC power is the dominant emission component, the synchrotron emission is still comparable. At the highest accretion rates right below the thin disk threshold (LHAF regime), $\alpha_c < 1$ and the IC radiation becomes highly efficient. We give in Section~\ref{sssec:ADAFtemp} details of how the temperature $\theta_e$ is treated for each regime.

The third component of the ADAF emission spectrum comes from bremsstrahlung of the thermal electrons, which provides a flat spectrum with an exponential drop at $T_e$
\begin{equation}
\begin{split}
\label{eq:bremslum}
    L_{\nu,\textrm{brems}} = 1.83 \times 10^{17} \left(\frac{\alpha}{0.1}\right)^{-2} \left(\frac{c_1}{0.5}\right)^{-2} \ln{\left(\frac{r_\textrm{max}}{r_\textrm{min}}\right)} F(\theta_e) \\ \left(\frac{T_e}{5\times10^{9}~\textrm{K}}\right)^{-1}e^{-(h\nu/kT_e)} m \dot{m}^2\textrm{ ergs} \textrm{ s}^{-1}\textrm{Hz}^{-1}~,
\end{split}
\end{equation}
where $r_\textrm{max}$ is $\mathcal{O}(10^3)$ and where $F(\theta_e) \sim \mathcal{O}(1)$ is given by~\cite{Mahadevan:1996jf}
\begin{align}
\begin{split}
\label{eq:ftheta}
        F(\theta_e)= 
    \begin{cases}
    4\left(\dfrac{2\theta_e}{\pi^3}\right)^\frac{1}{2} (1+1.78 \theta_e^{1.34}) + 1.73\theta_e^\frac{3}{2}(1+1.1\theta_e+\theta_e^2-1.25\theta_e^\frac{5}{2}),& \theta_e \leq 1~,\\
    \left(\dfrac{9\theta_e}{2\pi}\right)\left(\ln{[1.12\theta_e+0.48]}+1.5\right)+2.30\theta_e(\ln{[1.12\theta_e]} + 1.28),& \theta_e \geq 1~.
\end{cases}
\end{split}
\end{align}
While bremsstrahlung produces the majority of hard X-rays and gamma rays, IC typically generates more power in the relevant range of high hydrogen optical depth $10\textrm{ eV}$ to $100\textrm{ eV}$. Since the total bremsstrahlung power is also negligible relative to the combined synchrotron and IC power in the relevant parameter space, it will thus not contribute significantly to gas heating. For reference, a typical ADAF spectrum can be found in Fig. 1 of Ref.~\cite{2003ApJ...598..301Y}.

\subsubsection{ADAF - temperature considerations}
\label{sssec:ADAFtemp} 

For ADAF, the electron temperature $\theta_e$ is determined by balancing the heating and radiation processes~\cite{Mahadevan:1996jf}. At low $\dot{m}$, direct viscous electron heating $\delta q_+$ is dominant, and is superseded by ion-electron collisional heating $q_{ie}$ at high accretion rates. This results in different temperature considerations for each of the ADAF sub-regimes.

The heating contributions are given by
\begin{align}
\begin{split}
\label{eq:adafheating}
    \delta q_+ ~=&~   9.39\times10^{38}~\dfrac{\textrm{erg}}{\textrm{s}}~ \delta\left(\frac{1-\beta}{f}\right)c_3 m\dot{m}r_\textrm{min}^{-1} \\
    q_{ie} ~=&~ 1.2\times10^{38}~\dfrac{\textrm{erg}}{\textrm{s}}~g(\theta_e)\alpha^{-2}c_1^{-2}c_3 \beta m \dot{m}^2 r_\textrm{min}^{-1}~,
\end{split}
\end{align}
where $f$ is the fraction of energy advected that is $\sim 1$ for ADAF flows, $\delta = 0.3$ is the fraction of the viscously dissipated energy that directly heats
electrons, $\beta = 10/11$ is the ratio of gas pressure to total pressure, $r_\textrm{min}=3r_s$ is the minimum flow radius set by the ISCO radius, $\alpha = 0.1$ is the viscosity parameter and constants $c_1=0.5$, $c_3=0.3$ in the formalism of Ref.~\cite{Mahadevan:1996jf} characterize the ADAF solution~\cite{Narayan:1994xi} as before. Here, $g(\theta_e)$ is an $\mathcal{O}(1)$
function
\begin{equation}
\label{eq:gtheta}
    g(\theta_e) = \frac{1}{K_2(1/\theta_e)}\left(2+2\theta_e + \frac{1}{\theta_e}\right) e^{-(1/\theta_e)} 
\end{equation}
 that takes on the values of $4.5 - 0.63$ in the temperature range $2\times10^{9} \leq T_e \leq 10^{10}$.
For computational simplicity, we use the approximation $g(\theta_e) \simeq 1.19 ~\theta_e^{-1.15}$, which has a maximum error of $\sim 20\%$~\cite{Mahadevan:1996jf} over our temperature range of interest. When the two electron heating processes are equal, $\delta q_+ = q_{ie}$, the accretion rate is
\begin{equation}
\label{eq:heatingequality}
    \dot{m}_\textrm{eq} = 2.0\times10^{-3} \left(\frac{\alpha}{0.1}\right)^2\left(\frac{\delta}{0.1}\right)\left(\frac{1-\beta}{\beta}\right)\left(\frac{c_1}{0.5}\right)^2 \left(\frac{f}{1.0}\right)^{-1} g(\theta_e)^{-1}~.
\end{equation}
This is similar to our condition $\dot{M}_{\textrm{ADAF}}=10^{-3}(\alpha/0.1)^2$ of Sec.~\ref{sssec:ADAF}. Thus, in the ``eADAF" and ``ADAF" regimes, we use $\delta q_+$ heating and $q_{ie}$ heating for ``LHAF".

For $\dot{m}<\dot{M}_\textrm{ADAF}$, direct viscous heating balances the synchrotron and IC emission, which are of comparable magnitude. The temperature can be estimated as~\cite{Mahadevan:1996jf}
\begin{equation}
\label{eq:lowmdottemp}
    \theta_e \simeq 0.2{A_c^{-\frac{1}{7}}}\delta^\frac{1}{7}\alpha^\frac{3}{14}(1-\beta)^{-\frac{1}{14}}r_\textrm{min}^{\frac{3}{28}}\left(\frac{M}{M_\odot}\right)^\frac{1}{14}\dot{m}^{-\frac{5}{28}}~.
\end{equation}
Here $A_c$ is prefactor determined by the relative contributions of synchrotron and IC cooling in the total cooling rate, ranging from $0.95$ to $1.4$. As $\dot{m}$ increases, the accretion flow radiates more efficiently through IC scattering, cooling the electron plasma. We approximate $A_c^\frac{1}{7} = 1.1$ for $\dot{m}<\dot{M}_\textrm{eADAF}$ , and $A_c^\frac{1}{7} = 1.3$ for $\dot{M}_\textrm{eADAF}<\dot{m}<\dot{M}_\textrm{ADAF}$.

In the LHAF regime, the dominant IC emission is highly sensitive to temperature variations and balanced by ion electron $q_{ie}$ heating. Instead of directly evaluating this, we employ a more efficient iterative solution scheme modified from that developed in Ref.~\cite{Mahadevan:1996jf} to self-consistently determine the resulting temperature\footnote{Eq. (45) in Ref.~\cite{Mahadevan:1996jf} is used in lieu of Eq. (46) (which contains a sign error), the approximations for $x_M \simeq 4\times 10^3 \dot{m}^{1/4}$ and $g(\theta_e)$ are used in Eq. (47), and the new and old values of the temperature are averaged to improve convergence.}. Starting with an initial guess of $\alpha_c = 0.75$, Eq.~\eqref{eq:ICspectrum} can be solved for the temperature,
\begin{equation}
\label{eq:tempiter}
    \theta_e = \frac{\Big(\sqrt{4\tau_{es}^{-1/\alpha_c}-3}\Big)-1}{8}~.
\end{equation}
This result is then used to solve iteratively for a new $\alpha_c$ in Eq.~\eqref{eq:alphac} as
\begin{equation}
\label{eq:alphaiter}
    \alpha_{c,\textrm{new}} = 1-\frac{\ln\left[(1-\alpha_{c,\textrm{old}})\dfrac{q_{ie}}{\nu_p L_{\nu_p}}+1\right]}{\ln C_f}~,
\end{equation}
where
\begin{align}
\begin{split}
\label{eq:alphaiter2}
    \frac{q_{ie}}{\nu_p L_{\nu_p}} ~=&~ 1.78\times10^{-5}\theta_e^{-8.15}\alpha^{-\frac{1}{2}}\beta(1-\beta)^{-\frac{3}{2}}r_\textrm{min}^{\frac{3}{4}}\left(\frac{M}{M_\odot}\right)^\frac{1}{2}\dot{m}^{-\frac{1}{4}}~,\\
    C_f ~=&~ 20.2\alpha^{\frac{1}{2}}(1-\beta)^{-\frac{1}{2}}r_\textrm{min}^{\frac{5}{4}}\theta_e^{-1}\left(\frac{M}{M_\odot}\right)^{\frac{1}{2}}\dot{m}^{-\frac{3}{4}}~.
\end{split}
\end{align}
After each iteration, the new and old values of the temperature are averaged to speed up convergence.

\subsection{Dynamical friction}
\label{ssec:dynfric}

Dynamical friction due to gravitational interactions of traversing PBHs with the surrounding medium can both heat the gas and change the PBH velocity distribution. In contrast to photon emission and outflows, dynamical friction has the advantage of directly depositing kinetic energy, thus heating the gas without any additional considerations of attenuation length for emitted particles and stopping power. 

While some of the other heating mechanisms we have considered have a high degree of uncertainty, dynamical friction follows a straightforward description via the ``gravitational drag'' force given by~\cite{2008gady.book.....B}
\begin{align}
\begin{split}
\label{eq:dynfricgen}
    F_\textrm{dyn} =& -\frac{4\pi G^2 M^2 \rho}{v^2} I \\
    =& -3.95\times10^{12}\textrm{kg~m~s}^{-2} \left(\frac{M}{M_\odot}\right)^2 \left(\frac{\rho}{1\textrm{ GeV}\textrm{ cm}^{-3}}\right)\left(\frac{v}{10\textrm{ km}\textrm{ s}^{-1}}\right)^{-2} I~,
\end{split}
\end{align}
where $I$ is a velocity-dependent geometrical factor. For a collisionless medium, such as a particle DM candidate with negligible interactions, this factor takes the form
\begin{equation}
\label{eq:colllessI}
    I_\textrm{cl} = \ln\left(\frac{r_\textrm{max}}{r_\textrm{min}}\right)\left[\textrm{erf}\left(\frac{v}{\sqrt{2}\sigma_{\rm DM}}\right)-\frac{2v}{\sqrt{2\pi}\sigma_{\rm DM}}e^{-\frac{v^2}{2\sigma_{\rm DM}^2}}\right]~,
\end{equation}
where $\sigma_{\rm DM}$ is the DM velocity dispersion,  $r_\textrm{max}$ denotes the size of the surrounding affected system, which we identify with a gas cloud size and $r_\textrm{min}$ is the size of the perturbing body, which we take to be the Schwarzschild radius of the BH.

For the gaseous atomic hydrogen medium of Leo T, $I$ depends on the Mach number $\mathcal{M} = v/c_s$, where $c_s$ is the medium sound speed, and whether the motion of the perturbing PBH is subsonic ($\mathcal{M}<1$)  or supersonic ($\mathcal{M}>1$)~\cite{1999ApJ...513..252O}
\begin{align}
\begin{split}
\label{eq:gasI}
        I= 
    \begin{cases}
    \frac{1}{2}\ln\left(\frac{1+\mathcal{M}}{1-\mathcal{M}}\right)-\mathcal{M},& \mathcal{M}<1~,\\
    \frac{1}{2}\ln\left(1-\frac{1}{\mathcal{M}^2}\right)+\ln\left(\frac{r_\textrm{max}}{r_\textrm{min}}\right),& \mathcal{M}>1~.
    \end{cases}
\end{split}
\end{align}
The resulting output power due to dynamical friction for PBH traversing gas is also in this case the gas heating power $\mathcal{H}$ as the gas is directly heated, is given by
\begin{equation}
\label{eq:dynpowergen}
    \mathcal{H}_{\rm dyn} = P_\textrm{dyn} = F_{\rm dyn} v =  3.95\times10^{23} \textrm{ erg}\textrm{ s}^{-1} \left(\frac{M}{M_\odot}\right)^2\left(\frac{\rho}{1 \textrm{ GeV}\textrm{ cm}^{-3}}\right)\left(\frac{v}{10\textrm{ km}\textrm{ s}^{-1}}\right)^{-1} I~.
\end{equation}

We can estimate the change in PBH velocity due to dynamical friction using Eq.~\eqref{eq:colllessI}. With dwarf galaxies in mind, let us consider a gaseous system with DM density $\rho_{\rm DM}$ in the central region of the system that is significantly higher than the gas density $\rho = n m_p$. The effect of dynamical friction on the PBH velocity can be seen from comparing the total work done on the PBH traversing the system to its kinetic energy
\begin{equation}
\label{eq:dynfricratio}
\frac{F_\textrm{dyn}r_\textrm{max}}{\frac{1}{2}Mv^2} =-6.13\times10^{-7}\left(\frac{M}{M_\odot}\right)\left(\frac{\rho}{1\textrm{ GeV}\textrm{ cm}^{-3}}\right)\left(\frac{v}{10\textrm{ km}\textrm{ s}^{-1}}\right)^{-4}\left(\frac{r_\textrm{max}}{500\textrm{ pc}}\right) I~.
\end{equation}
The effect on the PBH velocity is insignificant throughout most of the parameter space of interest, unless we consider gaseous environments of significantly higher density (in either gas or DM) or super-massive PBHs with mass $M\gtrsim10^5 M_\odot$. We therefore ignore dynamical friction effects on PBH velocity. We note that lower relative velocities will generally increase the PBH accretion rate, resulting in an increase of the photon and outflow power delivered to the gaseous system. Hence, ignoring the velocity modification yields more conservative bounds on PBH gas heating.

\subsection{Accretion mass outflows/winds}
\label{ssec:outflows}

Mass outflows (winds) composed primarily of protons are expected to be significant for hot ADAF flows, contributing to the heating of the surrounding medium~\cite{Yuan:2014gma}. Unlike the highly beamed jets\footnote{As jets are typically associated with Kerr black holes, they would require a separate treatment and we do not consider them here.}, outflows are not highly relativistic and cover wider angular distributions. However, there is significant uncertainty in the description of outflows.

The outflows reduce the in-falling matter at smaller radii and the accretion rate in the presence of outflows can be approximately modelled by a self-similar power-law~\cite{Blandford:1998qn}
\begin{equation}
\label{eq:inflowgen}
    \dot{M}_\textrm{in}(r) = \dot{M}_\textrm{in}(r_\textrm{out})\left(\frac{r}{r_\textrm{out}}\right)^s~,
\end{equation}
where the index $0 \leq s < 1$ is limited by energy and mass conservation. Here, $\dot{M}_\textrm{in}(r_\textrm{out})$ is the Bondi-Hoyle-Lyttleton accretion rate of Eq.~\eqref{eq:bondihoyle}. The corresponding outflow (wind) rate is~\cite{2006PASJ...58..965T}
\begin{equation}
\label{eq:outflowgen}
    \dot{M}_w(r) = r\frac{d\dot{M}_\textrm{in}}{dr} = s\dot{M}_\textrm{in}(r_\textrm{out})\left(\frac{r}{r_\textrm{out}}\right)^s~.
\end{equation}
From simulations, the exponent $s$ is estimated to be $0.4-0.8$~\cite{2012ApJ...761..130Y,Yuan:2014gma}.

A wide range of values has been considered in the literature for the outer radius $r_\textrm{out}$, including $100 r_s$~\cite{Xie:2012rs}, $500 r_s$~\cite{2012ApJ...761..130Y},  $r_B$~\cite{2011MNRAS.415.1228P}, to $10 r_B$~\cite{2013ApJ...767..105L}. The resulting outgoing wind has a velocity that is a fraction $f_k\simeq0.1-0.2$~\cite{2013ApJ...767..105L,2013IAUS..290...86Y,2012ApJ...761..130Y} of the Keplerian velocity $v_k$ at the radius at which it is ejected (due to $Be/v_k^2$, where $Be$ is the Bernoulli parameter, being roughly constant~\cite{2012ApJ...761..130Y}), i.e. $v(r) \simeq f_k \sqrt{G M/r}$. Most of the kinetic energy of the wind comes from the inner region, with the kinetic energy per particle determined at the ISCO radius with $v_{\rm in} = v(r_{\rm in} = 3 r_s)$ being
\begin{equation}
E_\textrm{kin} \simeq \frac{1}{2} m_p v_{\rm in}^2 = 782\textrm{ keV} \left(\dfrac{r_\textrm{in}}{3r_s}\right)^{-1}\left(\dfrac{f_k}{0.1}\right)^2~,
\end{equation}
which is independent of PBH mass.

We derive the energy distribution of the wind, $f(E)$, and integrate it with the energy deposited per proton $\Delta(E)$. The heat deposited is limited by the value of the energy loss per unit length $x$ stopping power $dE/dx = nS(E)$, which we take from Ref.~\cite{2019PhRvA..99d2701B} (see their Fig.~9),
\begin{equation}
\label{eq:energydepo}
    \Delta E = \int \frac{dE}{dx} dx \simeq \textrm{min}(E,n S(E) r_\textrm{sys})~.
\end{equation}
The total heat deposited in the cloud is then
\begin{equation}
\label{eq:outflowint}
    \mathcal{H}_\textrm{outflow} = \int_{r_\textrm{in}}^{r_\textrm{out}} f_h \Delta E \frac{\mathcal{D}}{\mu m_p}\frac{d\dot{M}_\textrm{in}}{dr} dr = \int_{E(r_\textrm{min})}^{E(r_\textrm{out})} f_h \Delta E f(E) dE~, 
\end{equation}
where we estimate the fraction of energy deposited as heat to be $f_h\sim1/3$ and
where the distribution function is given by
\begin{align}
\begin{split}
    \label{eq:fe}
    f(E) ~=&~ s\dot{M}_\textrm{in}(r_\textrm{out})\left(\frac{r}{r_\textrm{out}}\right)^{s}\frac{1}{E} \\
    ~=&~ 3.54\times10^{22}~\frac{\textrm{ erg}}{\textrm{ eV}^{2}\textrm{ sec}}~ s(2.35\times10^{8})^s \left(\frac{M}{M_\odot}\right)^2 \left(\frac{n}{1\textrm{ cm}^{-3}}\right)\\ 
    &~\times\left(\frac{\Tilde{v}}{10\textrm{ km}/\textrm{s}}\right)^{-3} \left(\frac{\mathcal{D}}{1}\right)\left(\frac{r_\textrm{out}}{r_s}\right)^{-s}f_v^{2s}E^{-s-1}~.
\end{split}
\end{align}
We have chosen a duty cycle factor of $\mathcal{D}= 1$, as discussed below. 

Magnetic fields in the ISM can trap the protons from the accretion disk wind, effectively increasing the stopping power and the heat deposition. Since the strength, orientation, and distribution of magnetic fields in Leo T are not known\footnote{See related discussion for positron propagation, which even in the case of Milky Way is highly uncertain~\cite{Panther:2018xvc}.}, we conservatively omit any such factors.
 
\subsubsection{Duty cycle}
\label{sec:duty}

The wind outflow feedback can affect the PBH accretion and hence the emission efficiency. If the outflow feedback is very strong, heated ISM at the Bondi radius will be blown away, resulting in reduced accretion. On the other hand, if accretion is terminated so will be the wind and feedback. 
Wind duty cycle can be estimated following~Ref.~\cite{Ioka:2016bil} by comparing the dynamical and accretion timescales. If the accretion timescale is significantly longer than the dynamical timescale, we do not expect accretion to be efficient.

\begin{figure}[tb]
\begin{center}
\includegraphics[width=.5\textwidth]{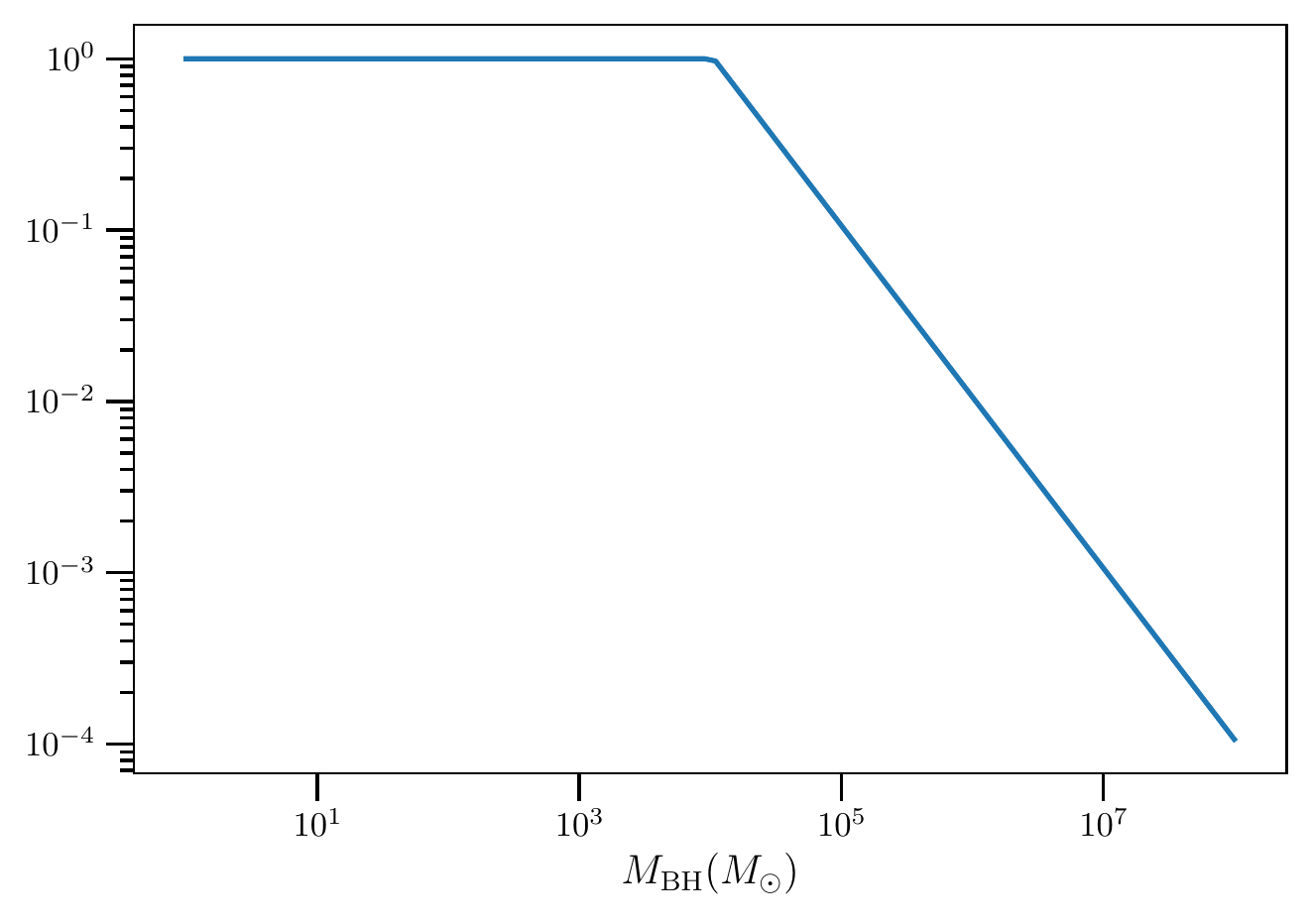}
\caption{PBH accretion wind outflow duty cycle $\mathcal{D}$, for characteristic $v = 10$~km/s and system radius $r_{\rm sys} = 1$~pc.}\label{fig:dutyc}
\end{center}
\end{figure}

The dynamical time of the system (i.e., the crossing time) is defined as
\begin{equation}
   t_{\rm dyn} = \frac{r_{\rm sys}}{v},
\end{equation}
while the accretion time scale (i.e., free-fall time scale) is defined as
\begin{equation}
	t_{\rm ff} = \frac{r_B}{v_{\rm ff}}=\frac{GM}{\sqrt{2}v^3}~,
\end{equation}
where $v_{\rm ff} = \sqrt{2GM/r_B}$ is the free-fall velocity, and $r_B=GM/v^2$ is the Bondi radius as before. Then we can consider the duty cycle $\mathcal{D}$ to be
\begin{equation}
\mathcal{D} = \max\left[1, \frac{t_{\rm dyn}}{t_{\rm ff}}\right]	= \max\left[1, \frac{\sqrt{2}r_{\rm sys}v^2}{GM}\right]~.
\end{equation}
We plot the duty cycle $\mathcal{D}$ on Fig.~\ref{fig:dutyc}, assuming characteristic $v=10$~km/s (here we take $\tilde{v} \sim v$) and $r_{\rm sys}=1$~pc values. For $M\lesssim 10^{4} M_\odot$, we estimate the factor to be 1, i.e. the duty cycle has no significant effect on Eq.~\eqref{eq:fe}. Above $M \gtrsim 10^4 M_{\odot}$, we find PBHs to be constrained by the incredulity limit (see Eq.~\eqref{eq:incredlimit}), which justifies us in taking $\mathcal{D}=1$ throughout.

\section{Astrophysical Systems}

We now apply our analysis to two examples of astrophysical systems, namely Milky Way gas clouds and dwarf galaxies. Although we limit our analysis to these systems, the methods outlined here can be readily extended to other gas clouds.  Requiring the thermal balance of the gas cloud between standard cooling processes and heating from PBH, we constrain the PBH mass fraction $f_\textrm{PBH}$. A similar approach has been used to set limits on particle DM~\cite{Bhoonah:2018wmw,Farrar:2019qrv,Wadekar:2019mpc}. We find that for PBHs the different heating processes are particularly favorable to explore within the dwarf galaxy Leo T.

Natural heating sources (e.g. stellar radiation) are for simplicity neglected in this study, which results in a more conservative upper bound on PBH heating. Since our argument relies on maintaining thermal equilibrium, only gas systems that are expected to be stable on sufficiently long timescales $\tau_\textrm{sys}$ are used (see discussion in Ref.~\cite{Farrar:2019qrv}).  We thus require the timescale for which the gas system remains stable to be longer than the cooling timescale of the gas $\tau_\textrm{therm}$
\begin{equation}
\label{eq:timescales}
    \tau_\textrm{sys} \gg \tau_\textrm{therm} = \frac{3nkT}{2\dot{C}}~,
\end{equation}
where $k$ is the Boltzmann constant and $\dot{C}$ is the gas cooling rate per volume. 

Furthermore, the disk formation timescale needs to be smaller than the system crossing timescale. This is directly analogous to the duty cycle considerations of Section~\ref{sec:duty} and is satisfied for our systems of interest. In the case of Leo T, this is not strictly necessary, since PBHs may orbit about the center of the dwarf galaxy. On the other hand, this could affect Milky Way cloud considerations, as PBHs do not spend all of their time inside them.   

In principle a multitude of processes can participate in gas temperature exchange, and there exist detailed numerical analyses involving a full chemistry network~\cite{Smith:2016hsc}. We employ approximate results obtained in Ref.~\cite{Wadekar:2019mpc} for our gas systems of interest. The volumetric cooling rate (erg cm$^{-3}$ s$^{-1}$) of atomic hydrogen gas is given by
\begin{equation}
\label{eq:coolinggen}
    \dot{C} = n^2 10^{[\textrm{Fe/H}]}\Lambda(T)~,
\end{equation}
where [Fe/H]$ \equiv \log_{10}(n_{\rm Fe}/n_{\rm H})_{\rm gas} - \log_{10}(n_{\rm Fe}/n_{\rm H})_{\rm Sun}$ is the metallicity, and $\Lambda(T) \propto 10^{[{\rm Fe/H}]}$ is the cooling function which can be fitted numerically. Using the results from Ref.~\cite{Smith:2016hsc} library, Ref.~\cite{Wadekar:2019mpc} obtained $\Lambda(T) = 2.51\times10^{-28}T^{0.6}$~(erg cm$^3$ s$^{-1}$), valid in the range $300~\text{K} < T < 8000~\text{K}$.

The total heat deposited in the cloud by the PBH 
\begin{equation}
  H_{\rm tot} = N_\textrm{PBH} H(M)= f_\textrm{PBH}\dfrac{\rho_\textrm{DM}V H(M)}{M}
\end{equation}
is required to be less than the total cooling rate $\dot{C} V$, yielding the condition on the PBH mass fraction
\begin{equation}
\label{eq:genbound}
    f_\textrm{PBH} < f_\textrm{bound} =\frac{M\dot{C}}{\rho_\textrm{DM}H(M)}~.
\end{equation}
Here $H(M)$ is the average heat deposited by one PBH of mass $M$.

\subsection{Gas systems with bulk relative velocity with respect to the DM}
\label{app:relvel}

A gas system with a bulk relative velocity $\vec{v}_b$ with respect to the Galactic frame, and thus the PBH in the DM, can be treated as follows.
Without loss of generality, consider the frame in which the bulk velocity is rotated to the z direction. Then the one dimensional Maxwellian velocity distributions are
\begin{align} 
\begin{split}
\label{eq:1dmaxwellian}
    \begin{cases}
    \dfrac{df_{v,x}}{dv}(v_x)=\dfrac{df_{v,y}}{dv}(v_y)=\dfrac{1}{\sqrt{2\pi}\sigma_v}e^{-\frac{v_x^2}{2\sigma_v^2}}\\ \\
    \dfrac{df_{v,z}}{dv}(v_z) = \dfrac{1}{\sqrt{2\pi}\sigma_v}e^{-\frac{(v_z-v_b)^2}{2\sigma_v^2}}
    \end{cases}
\end{split}
\end{align}
Imposing the constraint $v^2 = \sqrt{v_x^2+v_y^2+v_z^2}$, the relative velocity distribution is given by the integral
\begin{align}
\label{eq:maxwellintegral}
    \dfrac{df_v}{dv}(v) ~=&~ \frac{1}{(2\pi)^{\frac{3}{2}}\sigma_v^3}\int_{-\infty}^{\infty}dv_x\int_{-\infty}^{\infty}dv_y\int_{-\infty}^{\infty}dv_z \delta(v^2-v_x^2-v_y^2-v_z^2)e^{-\frac{v_x^2+v_y^2+(v_z-v_b)^2}{2\sigma_v^2}} \\
    =&~ \frac{e^{-\frac{v^2+v_b^2}{2\sigma^2}}}{(2\pi)^{\frac{3}{2}}\sigma_v^3}\int_{-v}^{v}dv_x\int_{-\sqrt{v^2-v_x^2}}^{\sqrt{v^2-v_x^2}}dv_y \frac{v}{\sqrt{v^2-v_x^2-v_y^2}}  \left[e^{-\frac{v_b\sqrt{v^2-v_x^2-v_y^2}}{\sigma_v^2}} + e^{-\frac{v_b\sqrt{v^2-v_x^2-v_y^2}}{\sigma_v^2}}\right]~. \notag
\end{align}
Switching to polar velocity coordinates and simplifying,
\begin{align}
\begin{split}
\label{eq:modmaxwellian}
    \dfrac{df_v}{dv}(v) ~=&~ \frac{2\pi v}{(2\pi)^\frac{3}{2}\sigma_v^3}e^{-\frac{v^2+v_b^2}{2\sigma_v^2}}\int_0^{v}dv_r \frac{v_r}{\sqrt{v^2-v_r^2}}\left[e^{-\frac{v_b\sqrt{v^2-v_r^2}}{\sigma_v^2}}+e^{\frac{v_b\sqrt{v^2-v_r^2}}{\sigma_v^2}}\right] \\
    ~=&~ \frac{v}{\sqrt{2\pi}\sigma v_b}\left[e^{-\frac{(v-v_b)^2}{2\sigma^2}}-e^{-\frac{(v+v_b)^2}{2\sigma^2}}\right]~.
\end{split}
\end{align}
In the limit $v_b\rightarrow 0$, Eq.~\eqref{eq:modmaxwellian} reproduces the standard Maxwellian distribution Eq.~\eqref{eq:maxwellian}.

\subsection{Milky-Way gas clouds}

\begin{figure}[tb]
\begin{center}
\includegraphics[trim={5mm 0mm 40 0},clip,width=.475\textwidth]{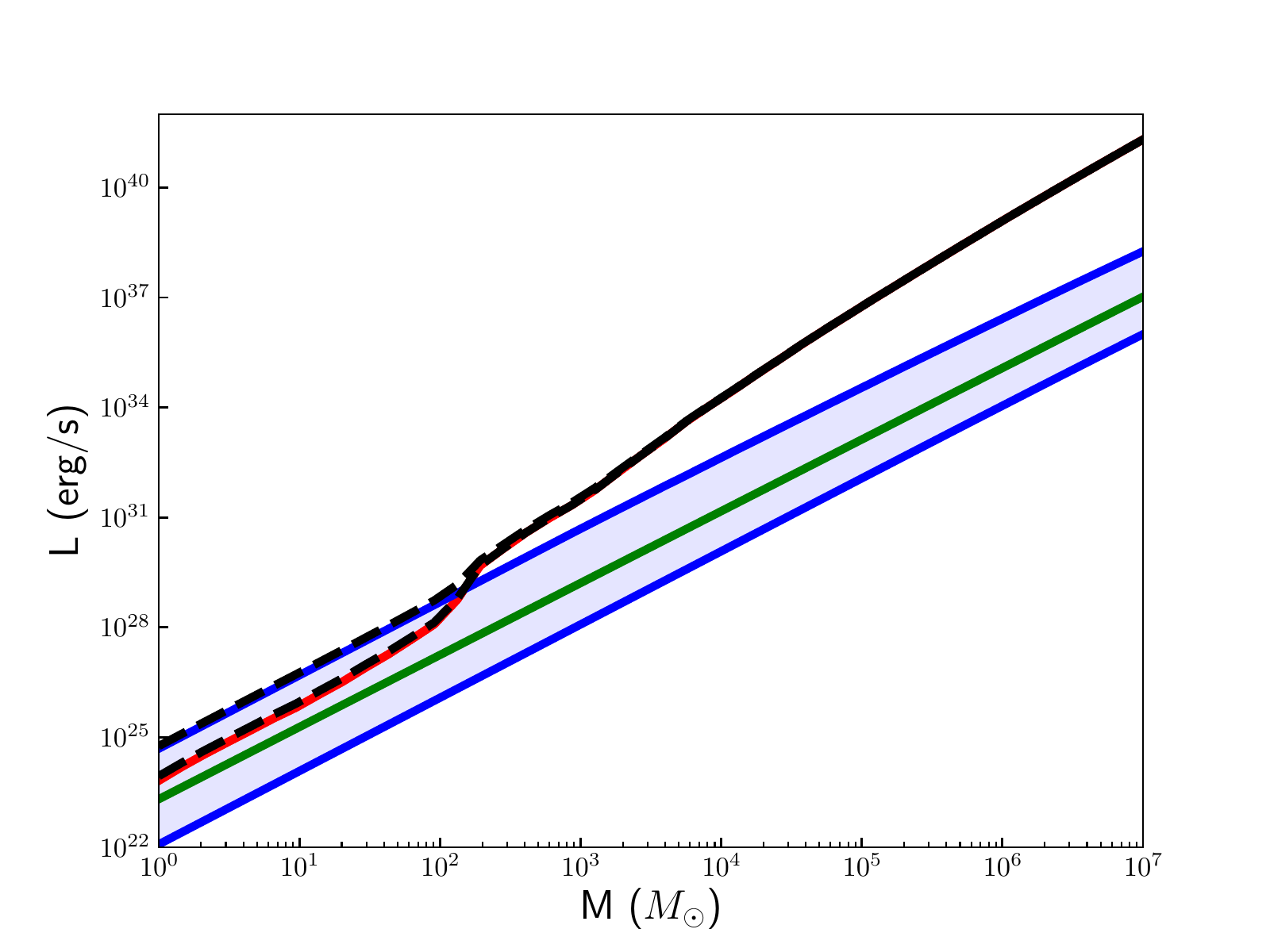}
\includegraphics[trim={5mm 0mm 40 0},clip,width=.475\textwidth]{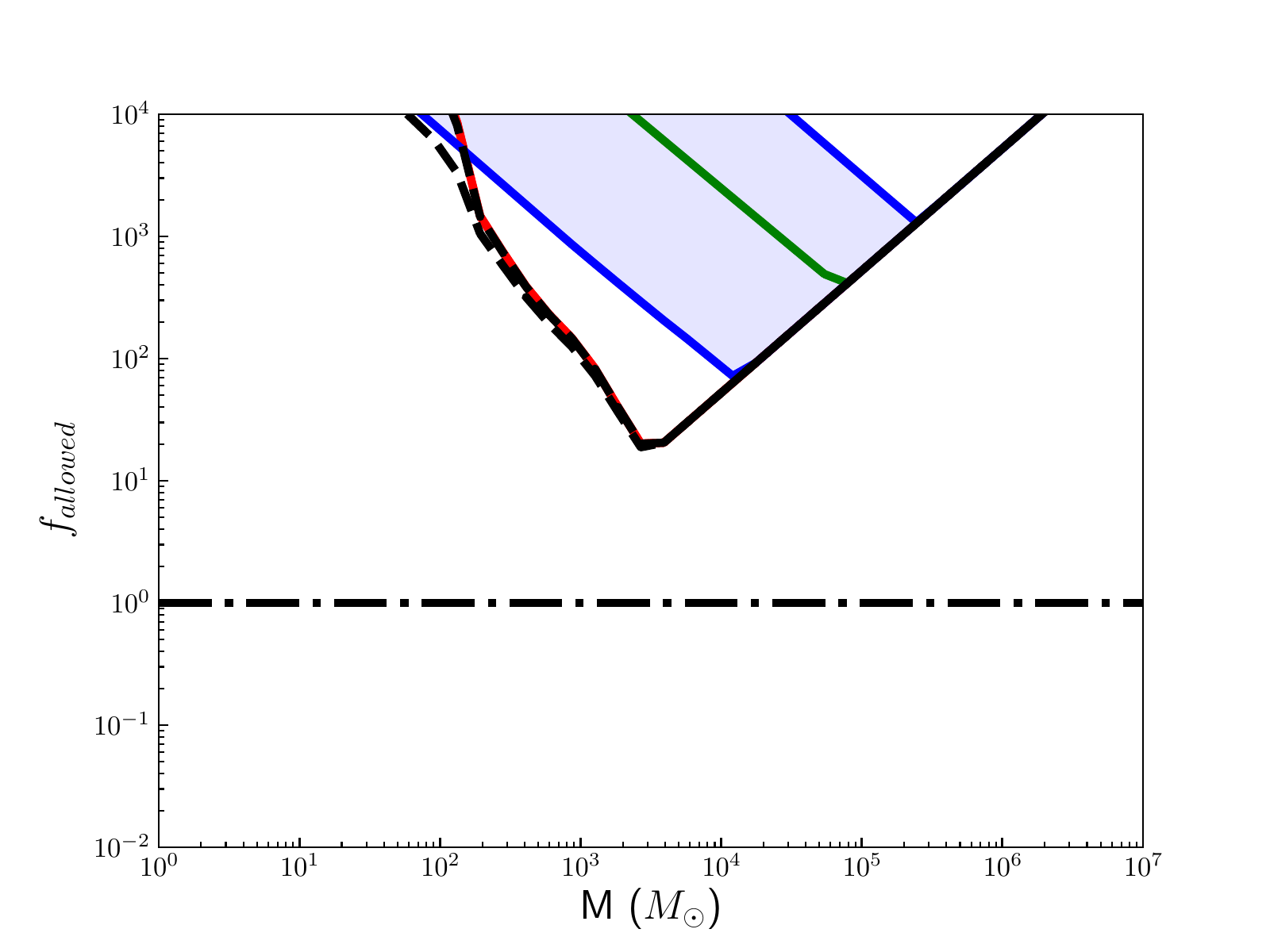}

\caption{  \label{fig:G33480graphs}  \textbf{Left:} 
Amount of heat absorbed by the Milky-Way gas cloud G33.4-8.0 from a single PBH of mass $M$. \textbf{Right:} Constraints from G33.4-8.0 on the DM fraction in PBHs. The reach of constraints is bounded by the positive slope solid black line indicating the incredulity limit. In both panels three heating components are shown: photon emission (red), dynamical friction (green), mass outflows (blue), as well as the total heat (black dashed). Both Model 1 and Model 2 of emission are shown, and the variation between them is shaded in blue. The observed line features are due to transitions between the eADAF, ADAF, LHAF, and thin disk regimes.}
\end{center}
\end{figure}

A variety of interstellar neutral hydrogen (HI) gas clouds exist in the inner Galaxy within a few hundred pc of the Galactic plane~\cite{2015ApJS..219...16P}.
In general, such Milky-Way clouds are not ideal candidates for studying PBHs due to the high relative rotational velocity, high velocity dispersion, and lower total DM mass (relative e.g. to dwarf galaxies) due to their smaller size. The last of these shortcomings limits the maximum mass considered for PBHs due to the incredulity limit of Eq.~\eqref{eq:incredlimit}. Thus, even if a cloud had favorable properties otherwise, it would not be able to provide significant constraints in the intermediate PBH mass range.
 
To demonstrate our analysis, we consider as an example a particular Milky-May cloud, G33.4-8.0. To describe G33.4-8.0, we use parameters in Refs.~\cite{2015ApJS..219...16P,2015ApJS..216...29B,2004ApJ...602..162H,2002ApJ...580L..47L,2010ApJ...722..367F,2001ApJ...559..318R,Wadekar:2019mpc}. Namely, the atomic hydrogen gas density is taken to be $n=0.4~\pm 0.1\textrm{ cm}^{-3}$, the cloud radius $r\simeq 30\textrm{ pc}$, the bulk relative velocity $v_{\rm bulk} = 220\textrm{ km/s}$, the temperature $T=400~\pm 90$~K, and the corresponding adiabatic sound speed $c_s = 2.4 \textrm{ km/s}$, and the cooling rate of $\simeq 2.1\times10^{-27} \textrm{ erg}\textrm{ cm}^{-3}\textrm{ s}^{-1}$. The DM density of the cloud is $\rho = 0.64\textrm{ GeV}/ \textrm{cm}^{3}$, assuming Navarro-Frenk-White (NFW) profile~\cite{Navarro:1995iw} for the Galactic DM halo, with the DM velocity dispersion of $124.4\textrm{ km/s}$. 

In Fig.~\ref{fig:G33480graphs}, we display the heating from a single PBH for G33.4-8.0, including contributions due to photon emission, dynamical friction and mass outflows. To obtain the limits in Fig.~\ref{fig:G33480graphs}, we numerically integrate Eq.~\eqref{eq:totalheateq}. To treat the PBH emission and its uncertainty, we consider two distinct input parameter sets\footnote{We do not consider a variation in PBH emission parameter $\delta$ between $0.1-0.5$ for our analysis, since it only corresponds to a $\mathcal{O}(\%)$ variation in the photon luminosity.} given in Table~\ref{tab:param}. The observed irregularities in the figure are due to transitions between the eADAF, ADAF, LHAF, and thin disk regimes. 

We also display in Fig.~\ref{fig:G33480graphs} the combined resulting constraints from G33.4-8.0 gas heating, bound by the incredulity limit 
\begin{equation}
    f_\textrm{bound} > 5.25\times10^{-4}\left(\frac{M}{M_\odot}\right)\left(\frac{r_\textrm{sys}}{30\textrm{ pc}}\right)^{-3}\left(\frac{\rho_\textrm{DM}}{0.64\textrm{ GeV}\textrm{ cm}^{-3}}\right)^{-1}~. 
\end{equation}
The largest PBH mass that can be bounded by G33.4-8.0 with this method is $1.9\times10^{3} M_\odot$. The weakness of Milky Way gas cloud bounds is due in large part to the high velocity dispersion as well as the co-rotation of the gas clouds in the Milky Way, which adversely affects accretion disk formation. In addition, the low metallicity of Leo T contributes to reduced cooling rates relative to G33.4-8.0 enhancing the effect of PBH heating.

\begin{table}[tb]
  \setlength{\extrarowheight}{2pt}
  \setlength{\tabcolsep}{5pt}
  \begin{center}
	\begin{tabular}{c|c|c|c|c|c|c|c}  \hline\hline
	   & $\delta$ & $\beta$ & $\alpha$ & $s$ & $r_{\rm outer}$ & $f_k$ & $\mathcal{D}$\\  
	\hline
	Model~1 & 0.3 & 10/11 & 0.1 & 0.5 & $r_B$ & 0.1 & 1\\  \hline
	Model~2 & 0.3 & 10/11 & 0.1 & 0.7 & $100r_s$ & 0.2 & 1\\   \hline\hline
	\end{tabular}
  \end{center}
\caption{\label{tab:param} Input parameters for PBH emission.}
\end{table}

\subsection{Dwarf galaxies}

Dwarf DM-rich galaxies provide a good setting to study DM-gas heating. In contrast to Milky-Way gas clouds, their lack of relative bulk velocity with respect to the DM as well as significant size that allows to weaken the incredulity limit makes them favorable systems for PBH heating. 

We will focus on Leo T, a transition type of galaxy between a dwarf irregular and a spheroidal, which is
a few hundred kpc away from the Milky Way. Leo T has been well studied and modeled theoretically and it has the desired properties, including low baryon velocity dispersion that is also favorable for PBH heating.

In the inner $r\lesssim 350\textrm{ pc}$ of Leo T, the gas is mostly composed of atomic hydrogen, but consists of highly ionized hydrogen outside this central region~\cite{2013ApJ...777..119F}. To minimize the cooling rate, we avoid the efficiently cooled ionized region (see e.g.~Ref.~\cite{Smith:2016hsc}) and focus our analysis on the central region. The atomic hydrogen number density ranges from a maximum of $\sim 0.2\textrm{ cm}^{-3}$ at the center decreasing to $0.03\textrm{ cm}^{-3}$ at the boundary, at $r=350\textrm{ pc}$~\cite{2013ApJ...777..119F}. We estimate the density to be a constant $0.07\textrm{ cm}^{-3}$ (roughly the average density in the central region), and justify this by noting that both the heating processes and cooling rate scale approximately as $n^2$. This average density of gas produces the column density of $n r_\textrm{sys} = 7.56\times10^{19}\textrm{ cm}^{-2}$, in the right range for Leo T.  Similarly, we assume a constant DM density of $1.75\textrm{ GeV}/\textrm{cm}^3$ which accounts for the total DM mass in Leo T of $10^{7} M_\odot$.

The hydrogen gas in the central region of Leo T is comprised both of a non-rotating warm component with velocity dispersion of $\sigma_g=6.9~ \textrm{km}/\textrm{s}$ and temperature $T\simeq 6000$~K~\cite{2008MNRAS.384..535R,2013ApJ...777..119F} as well as a subdominant cold component. For simplicity, we treat the entire hydrogen density as warm gas which results in a more conservative bound. From the adiabatic formula, the sound speed is estimated to be $c_s\simeq 9\textrm{ km/s}$ for a $6000$ K gas. The DM velocity dispersion is expected to be the same as the gas dispersion, so that $\sigma_{v}=\sigma_g$. Using the above input parameters as well as metallicity of $\textrm{[Fe/H]} \simeq -2$~\cite{2008ApJ...685L..43K} and assuming that gas metallicity is related to stellar metallicity in Eq.~\eqref{eq:coolinggen}, the resulting Leo T's cooling rate is taken to be $\dot{C} = 2.28\times10^{-30}\textrm{ erg}\textrm{ cm}^{-3}\textrm{ s}^{-1}$.

\begin{figure}[tb]
\begin{center}
\includegraphics[trim={5mm 0mm 40 0},clip,width=.475\textwidth]{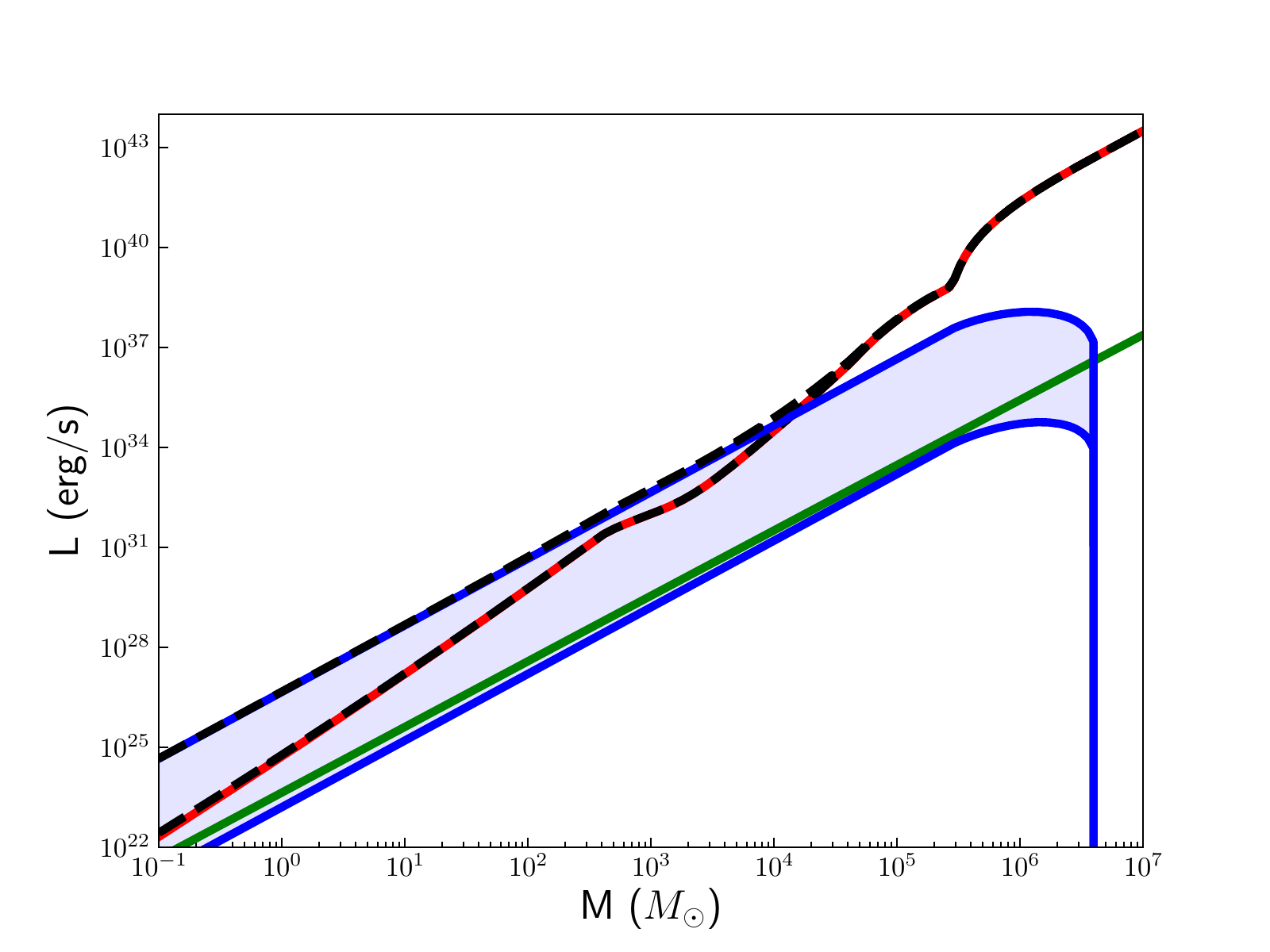}
\includegraphics[trim={5mm 0mm 40 0},clip,width=.475\textwidth]{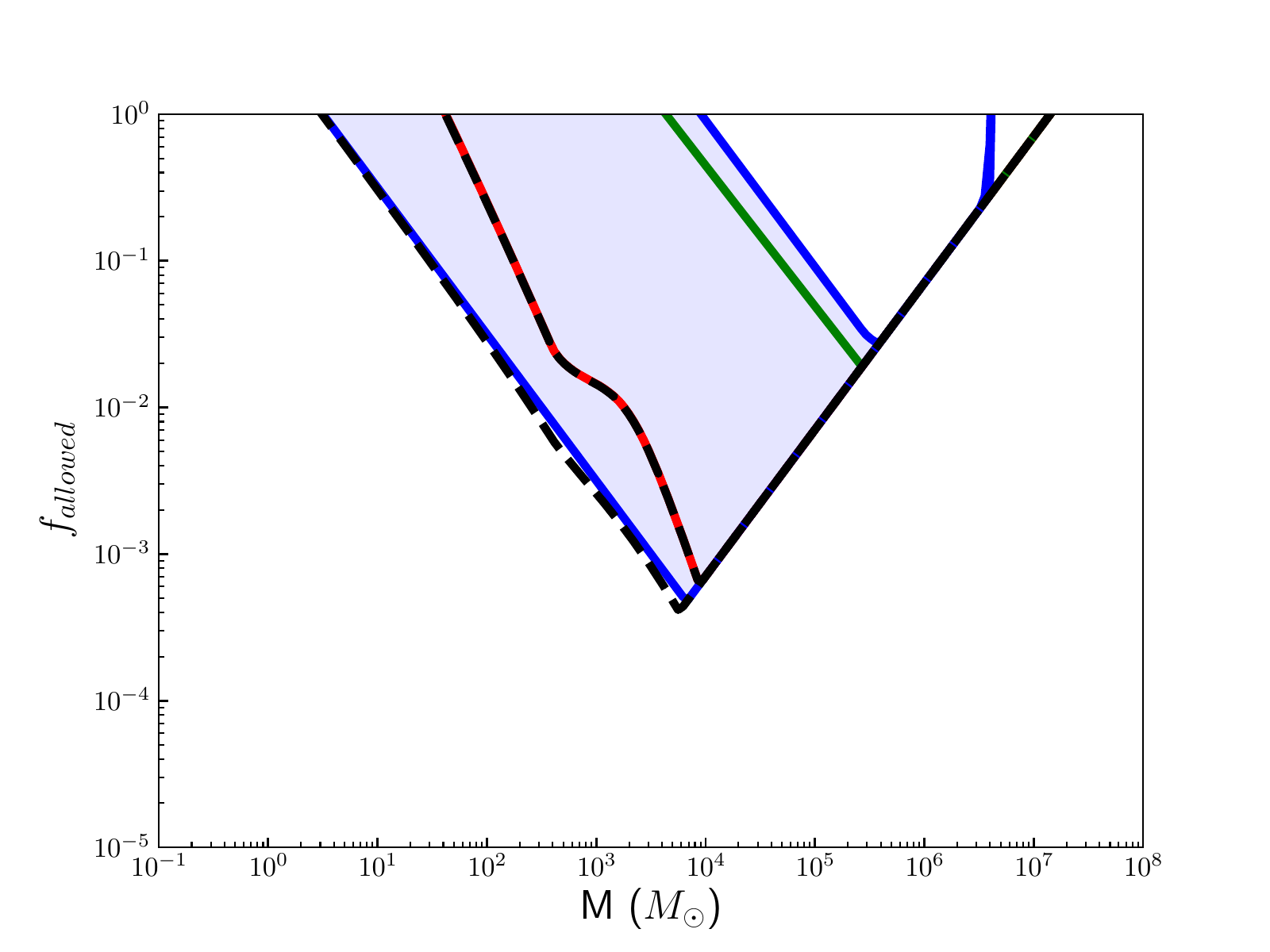}
\caption{  \label{fig:leotgraphs}  \textbf{Left:}
The amount of heat absorbed by Leo T dwarf galaxy from a single PBH of mass $M$. \textbf{Right:} Constraints from Leo T on the DM fraction in PBHs. The reach of constraints is bounded by the positive slope solid black line indicating the incredulity limit. The variation in the constraint from Models 1 and 2 are shaded in blue. The black line shows the combined constraint from Model 2. For both panels three heating components are shown: photon emission (red), dynamical friction (green), mass outflows (blue), as well as the total heat (black). The observed line features are due to transitions between the eADAF, ADAF, LHAF, and thin disk regimes. Both Model 1 and Model 2 of emission are shown, and the difference is shaded in blue.}
\end{center}
\end{figure}

In Fig.~\ref{fig:leotgraphs}, we display the heating from a single PBH for Leo T, including contributions due to photon emission, dynamical friction as well as mass outflows. As before, we numerically integrate Eq.~\eqref{eq:totalheateq} and consider PBH emission parameter variation as described in Tab.~\ref{tab:param}. We note that for the Leo T parameters, the condition for thin disks, $\dot{m}>0.07\alpha$, is satisfied only for very massive PBHs $M\gtrsim 10^{5}M_\odot$. 

We also display in Fig.~\ref{fig:leotgraphs} the combined resulting Leo T gas heating constraints, bound by the incredulity limit
\begin{equation}
    f_\textrm{bound} > 1.21\times10^{-7}\left(\frac{M}{M_\odot}\right)\left(\frac{r_\textrm{sys}}{350\textrm{ pc}}\right)^{-3}\left(\frac{\rho_\textrm{DM}}{1.75\textrm{ GeV}\textrm{ cm}^{-3}}\right)^{-1}~. 
\end{equation}
The largest PBH mass that can be bound by Leo T by this method is $8.26\times10^{6}M_\odot$. 

\begin{figure}[tb]
\begin{center}
\includegraphics[trim={5mm 0mm 40 0},clip,width=0.65\textwidth]{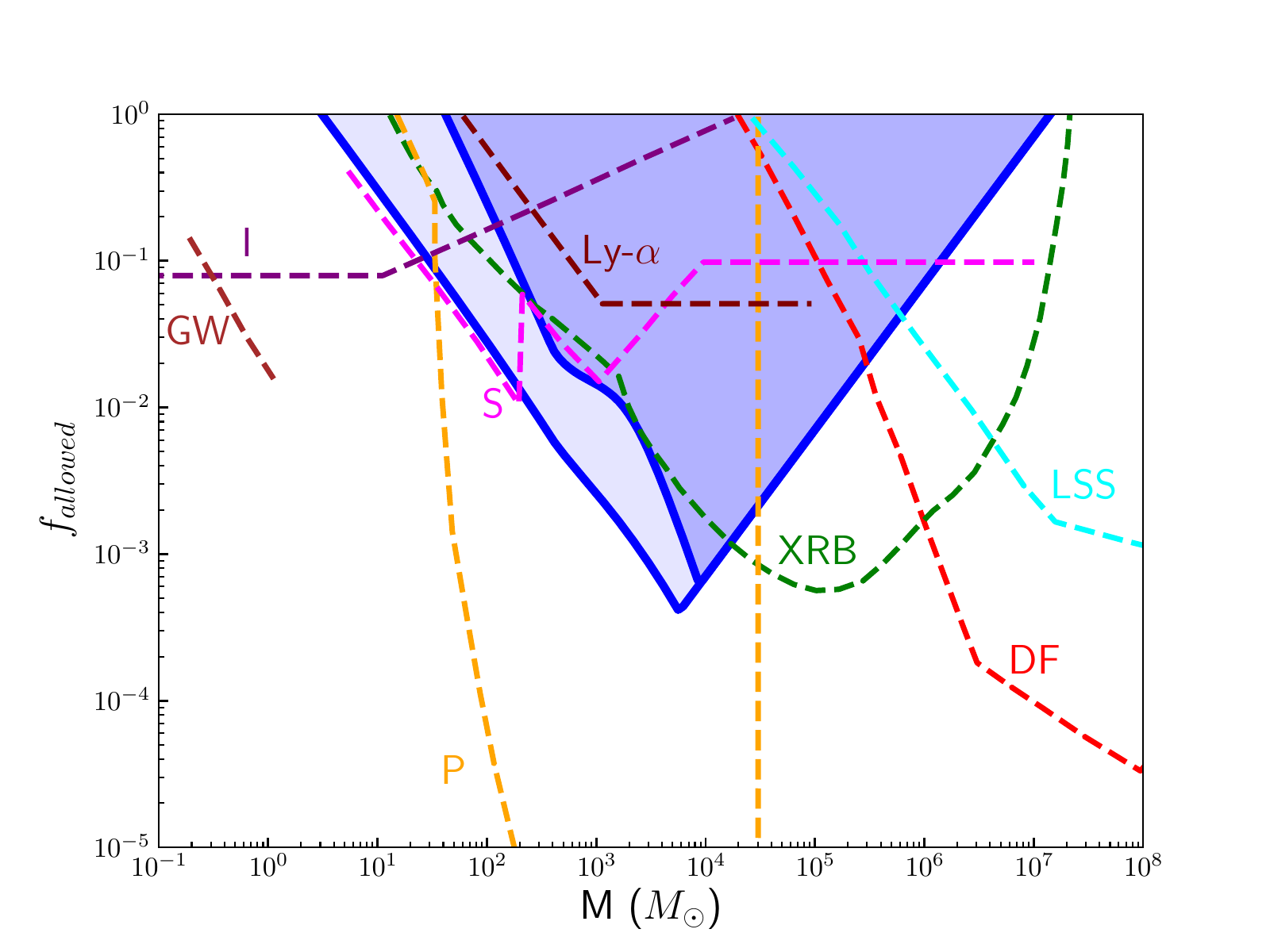}

\caption{  \label{leotfinalgraph} Constraints from Leo T dwarf galaxy on PBH gas heating are shown in blue. The light blue shaded band denotes variation due to PBH emission parameters. These constraints are bounded by the PBH incredulity limit (solid blue positive slope line). Other existing constraints (see Ref.~\citep{Carr:2020gox})  are shown by dashed lines, including Icarus~\cite{2018PhRvD..97b3518O}
(I) caustic crossing in purple, Planck~\cite{Ali-Haimoud:2016mbv,Serpico:2020ehh}
(P) in yellow, X-ray binaries~\cite{Inoue:2017csr}
(XRB) in green, dynamical friction of halo objects
(DF) in red, Lyman-$\alpha$~\cite{2019PhRvL.123g1102M}
(Ly-$\alpha$) in maroon, combined bounds from the survival of astrophysical systems in Eridanus II~\cite{Brandt:2016aco},
 Segue 1~\cite{2017PhRvL.119d1102K},
 and disruption of wide binaries~\cite{2014ApJ...790..159M}
(S) shown in magenta, large scale structure~\cite{Carr:2018rid}
(LSS) in cyan, and gravitational waves~\cite{LIGOScientific:2019kan}
(GW) in brown.}
\end{center}
\end{figure}

\section{Conclusions}
\label{sec:conclusions}

PBHs in the stellar as well as intermediate $\sim 10-10^3 ~M_{\odot}$ mass-range have garnished significant interest and could potentially be associated with LIGO gravity wave events. As PBHs traverse the ISM, they would interact with and heat the surrounding gas. We have considered several generic heating mechanisms, including accretion photon emission, dynamical friction as well as mass outflows/winds. The balance of system heating and cooling allows for a general test of PBHs in a cosmology-independent way. As a demonstration, we have applied our analysis to Milky-Way gas clouds and dwarf galaxies. Considering constraints on gas heating for a given gas system, we have derived new bounds from Leo T dwarf galaxy on PBHs contributing to DM across a broad mass-range of $M_{\rm PBH} \sim \mathcal{O}(1) M_{\odot}-10^7 M_{\odot}$. These limits do not depend on cosmology and establish a novel independent test of PBHs constituting part of the DM in the stellar and intermediate PBH mass range. Our analysis can be readily applied to other astrophysical systems of interest.

\acknowledgments
\addcontentsline{toc}{section}{Acknowledgments}
 
The work of G.B.G., A.K., V.T., and P.L. was supported by the U.S. Department of Energy (DOE) grant No. DE-SC0009937. This work was also supported by the MEXT
Grant-in-Aid for Scientific Research on Innovative Areas (No.~20H01895, No.~21K13909 for K.H.). A.K. was also supported by Japan Society for the Promotion of Science (JSPS) KAKENHI grant No.
JP20H05853. Y.I. is supported by JSPS KAKENHI grant No. JP18H05458, JP19K14772, program of Leading Initiative for Excellent Young Researchers, MEXT, Japan, and RIKEN iTHEMS Program. A.K., V.T., and Y.I. are also supported by the World Premier International Research Center Initiative (WPI), MEXT, Japan.

\appendix

\bibliography{PBH}
\addcontentsline{toc}{section}{Bibliography}
\bibliographystyle{JHEP}
\end{document}